\documentclass[%
 aip,
 amsmath,amssymb,
 reprint,%
]{revtex4-1}

\usepackage{graphicx}
\usepackage{dcolumn}
\usepackage{bm}

\begin{document}

\preprint{AIP/123-QED}

\title{Ab-initio calculations for electronic and optical properties of Er$_{\rm W}$ defects in tungsten disulfide}

\author{M. A. Khan}
\email{mahtabahmad.khan@fuuast.edu.pk}
 \affiliation{Department of Applied Physics, Federal Urdu University of Arts, Science and Technology, Islamabad, Pakistan}
\author{Michael N. Leuenberger}%
 \email{michael.leuenberger@ucf.edu}
\affiliation{ 
NanoScience Technology Center, Department of Physics, and College of Optics and Photonics, University
of Central Florida, Orlando, FL 32826, USA
}%

\date{\today}

\begin{abstract}
Ab-initio calculations for the electronic and optical properties of single-layer (SL) tungsten disulfide (WS$_2$) in the presence of substitutional Erbium defects (Er$_{\rm W}$) are presented, where the W atom is replaced by an Er atom. Defects usually play an important role in tailoring electronic and optical properties of semiconductors. We show that neutral Er defects lead to localized defect states (LDS) in the band structure due to the f-orbital states of Er, which in turn give rise to sharp transitions in in-plane and out-of-plane optical absorption spectra, $\alpha_{\parallel}$ and $\alpha_{\perp}$. We identify the optical transitions at 3 $\mu$m, 1.5 $\mu$m, 1.2 $\mu$m, 920 nm, 780 nm, 660 nm, and 550 nm  to originate from Er$_{\rm W}$ defect states. 
In order to provide a clear description of the optical absorption spectra, we use group theory to derive the optical selection rules between LDS for both $\alpha_{\parallel}$ and $\alpha_{\perp}$. 
\end{abstract}

\maketitle

\section{Introduction}

Single-layer (SL) transition metal dichalcogenides (TMDs) have attracted a lot of attention due to their intriguing electronic and optical properties, with a wide range of promising applications.\cite{review_TMDCS,review_TMDCS_2} SL TMDs are direct band gap semiconductors,\cite{Direct_Band_Gap_1,Direct_Band_gap_2} which can be used to produce smaller and more energy efficient devices, such as transistors and integrated circuits. Moreover, the band gap lies in the visible region which makes them highly responsive when exposed to visible light, a property with potential applications in optical detection. It is well-known that the exfoliation or growth processes can introduce defects and impurities in SL materials which can significantly alter their electronic, optical and magnetic properties.\cite{Sulfur_vacancies, Zhang_jap, Structural_Defects_Graphen}

Khan et al.\cite{Erementchouk2015,Khan2017} recently developed theoretical models based on density functional theory (DFT), tight-binding model, and 2D Dirac equation for the description of the electronic and optical properties of vacancy defects in TMDs, which are naturally occurring during different growth processes, such as mechanical exfoliation (ME), chemical vapor deposition (CVD), and physical vapor deposition (PVD).
 A central result of their papers is that group theory can be used to derive strict selection rules for the optical transitions, which are in excellent agreement with the susceptibility calculated by means of the Kubo-Greenwood formula using the Kohn-Sham orbitals. 
 
 Recently, Bai et al. have developed experimental methods to create Er-doped MoS$_2$ thin films using CVD growth\cite{Bai2016} and wafer-scale layered Yb/Er co-doped WSe$_2$ using pulsed laser deposition (PLD).\cite{Bai2018} One of the main motivations of doping TMDs with rare-earth ions (REIs) is that REIs inside an insulator or semiconductor crystal exhibit the unique property of having electrons in the unfilled 4f shell that is strongly isolated from crystal by the surrounding d shell. This property leads generally to high quantum yields, atom-like narrow bandwidths for optical transitions, long lifetimes, long decoherence times, high photostability, and large Stokes shifts. A prime example is Er-doped semiconductors and optical fibers that emit 1.5 $\mu$m light with ultra-narrow bandwidth and are therefore crucial for optoelectronic devices and optical telecommunication. Interestingly, Bai et al. show upconversion from 980 nm to 800 nm and simultaneous downconversion from 980 nm to 1550 nm using Er:MoS$_2$ atomic layers. In the case of Yb/Er:WSe$_2$ they observed downconversion from 980 nm to 1540 nm.  Some of the peaks in the theoretically calculated optical spectrum (see Fig.~\ref{fig:OS_Er_W}) are in good agreement with the available experimental data, in particular with the optical transition at 1.5 $\mu$m.\cite{Bai2016, Bai2018, YANG2007207}
 \begin{figure*}
	\begin{center}
		\includegraphics[width=7in]{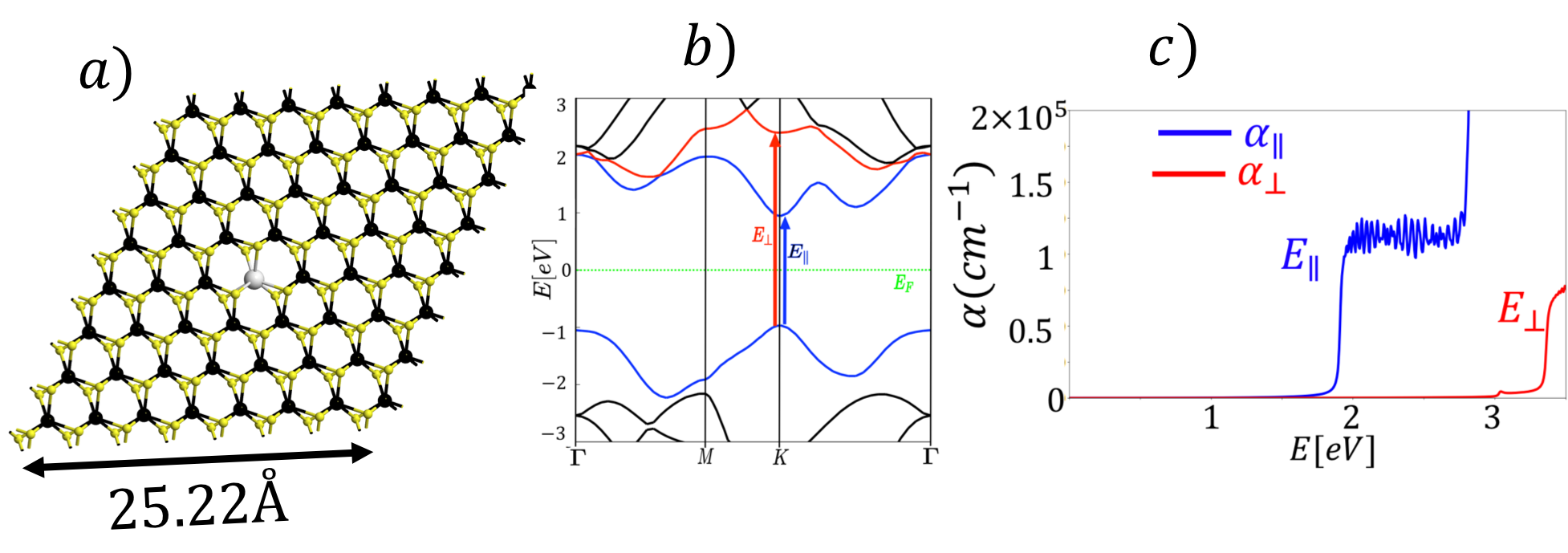}
	\end{center}
	\caption{a) Schematic diagram of $8\times8\times1$ supercell of WS$_2$ with Er$_{\rm W}$ defect, black (yellow) circles represent W (S) atoms, while grey circle represents Er atom. b) the band structure of the pristine WS$_2$ showing the in-plane $E_{\parallel}=1.9$ eV  and out-of-plane $E_{\perp}=3.4$ eV band gaps determined by the transitions $T_{\parallel}$ and $T_{\perp}$, respectively. c). Optical response of the pristine WS$_{2}$.  }
	\label{fig:Er_W_defect}
\end{figure*} 

 Pristine TMDs are invariant with respect to the reflection $\sigma_{h}$ about the Mo or W plane of atoms ($z=0$ plane). Therefore, electron states can be classified into two catagories: even and odd or symmetric and antisymmetric with respect to the $z=0$ plane. Khan et al. found that the even and odd bands in TMDs have two different band gaps $E_{g\parallel}$ and $E_{g\perp}$, respectively.\cite{Erementchouk2015,Khan2017} $E_{g\perp}$ has been usually neglected for pristine TMDs because of its substantially larger value and weak optical response as compared with $E_{g\parallel}$. Earlier studies\cite{Erementchouk2015,Khan2017,BG_Tune_MoS_2} show that the presence of VDs gives rise to LDS  in addition to the normal extended states present in conduction or valence bands in SL MoS$_{2}$. These LDS appear within the band gap region and they can also be present deep inside the valence band depending on the type of VD. Optical transitions between LDS across Fermi level appear as resonance peaks, both in $\alpha_{\parallel}$  and $\alpha_{\perp}$, which shows that odd states are necessary for understanding the properties of VDs in SL MoS$_{2}$.\cite{Erementchouk2015,Khan2017} The same symmetry considerations apply to Er:MX$_2$ when Er substitutes the M atom. 
 
 The goal of this paper is to demonstrate the existence of localized Er states inside the bandgap of WS$_2$, the ultra-narrow optical transitions due to the atom-like f-orbital states of Er, and the strict optical selection rules by means of a combination of ab-initio DFT calculations, the Kubo-Greenwood formula, and group theory.
\begin{figure*}[hbt]
	\begin{center}
		\includegraphics[width=7in]{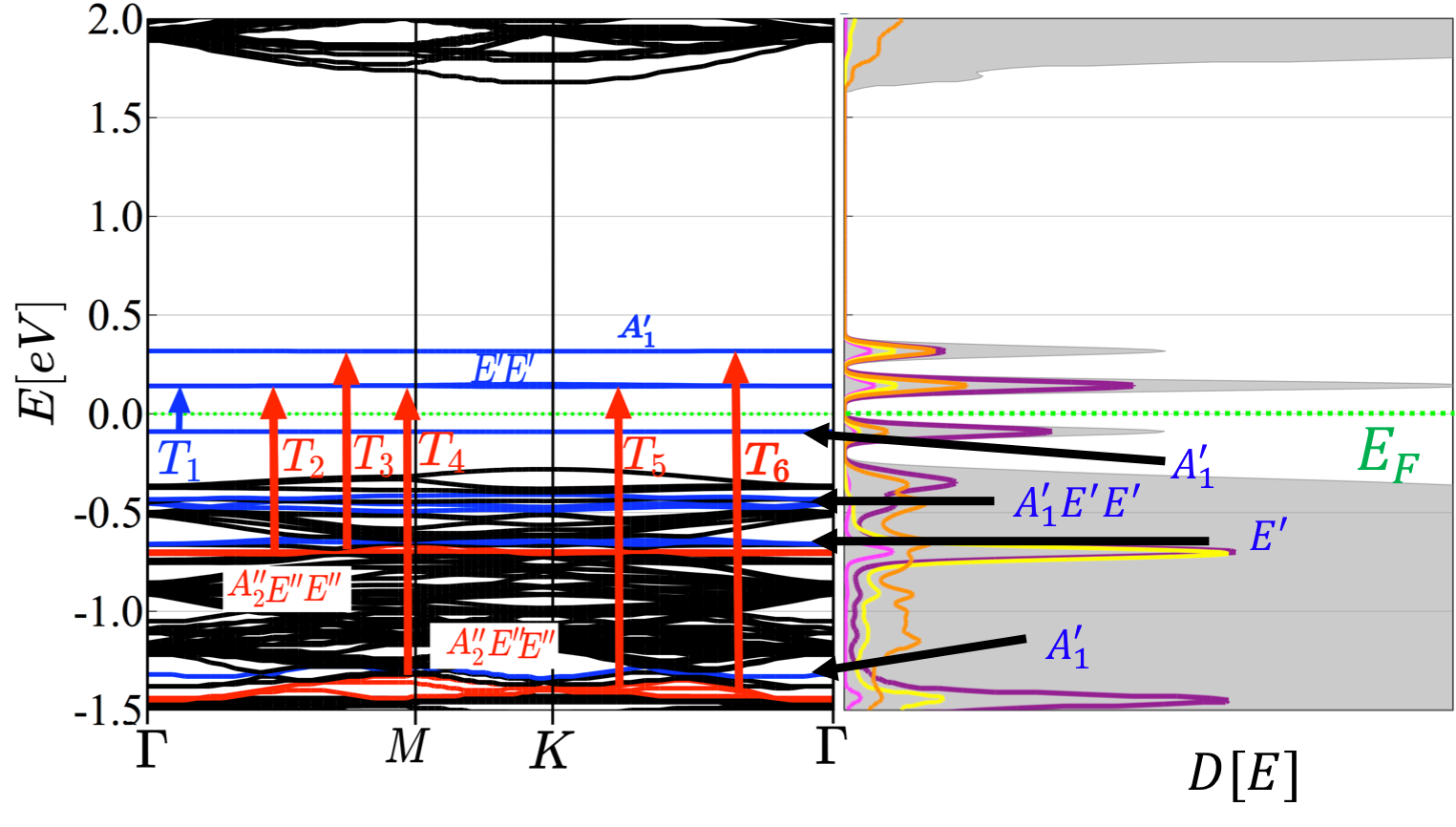}
	\end{center}
	\caption{Bandstructure and density of states, shaded grey region shows the total density of states and the colored curves shows the projected density of states (purple:$f-$orbitals of the Er atom, magenta:$d-$orbitals of Er atom, yellow:$p-$orbital of the neighboring S atoms and orange:$d-$orbitals of the next neighbor W atoms) of $8\times8\times1$ supercell of WS$_2$ containing an Er$_{\rm W}$ defect. The LDS are clearly visible as dispersionless (localized) states, some of which lie inside the bandgap, others lie inside the valence band of WS$_2$. The eigenstates corresponding to the LDS transform according to the irreducible representations of the point symmetry group $D_{\rm 3h}$. Vertical arrows indicate optical transitions corresponding to resonances shown in Fig. \ref{fig:OS_Er_W}.}
	\label{fig:bandstructure_Er_W}
\end{figure*}
 \section{Bandstructure}
 \label{sec:bandstructure}
 
 \subsection{Numerical Analysis}
 The model system consists of a periodic 2D superlattice of TMDs (Fig.~\ref{fig:Er_W_defect}a). All numerical calculations are carried out using density functional theory (DFT). The generalized gradient approximation (GGA) is used  with the perdew-burke-ernzerhof (PBE) parametrization\cite{PZ_functionals} of the correlation energy of a homogeneous electron gas calculated by Ceperley-Alder \cite{Ceperley_Alder}. The calculations are implemented within the Synopsis Atomistix Toolkit (ATK) 2019.12.\cite{QW_1} 
 The periodic structure of the superlattice allows one to characterize the electron states by the bandstructure $\epsilon_n(\mathbf{k})$, where $\mathbf{k}$ is the vector in the first Brillouin zone of the superlattice and $n$ enumerates different bands. We consider a $8 \times 8 \times 1$ (Fig.~\ref{fig:Er_W_defect}) supercell having $192$ number of atoms with an edge length of $25.22$ \AA. The Brillouin zone of the supercell is sampled by a $5 \times 5 \times 1$ $k$-mesh. In order to account for the correlation effects, a Hubbard U parameter of 2.5 eV is added for the $f$ electrons of Er. All the structures are geometrically optimized with a force tolerance of $0.05$ eV/\AA. 

Bandstructures are calculated along the $\Gamma$ $-$ $M$ $-$ $K$ $-$ $\Gamma$ path. Bandstructure of SL WS$_2$ for the pristine cases are plotted in Fig.~\ref{fig:Er_W_defect} 
 The results are in good agreement with previously reported values for band gap energy. 
 \cite{SOC_1,SOC_2,SOC_agreement_1,SOC_agreement_2} Fig.~\ref{fig:bandstructure_Er_W} shows the bandstructure of WS$_2$ in the presence of Er$_{\rm W}$ defects. Black lines denote regular electronic states within the valence or conduction bands while colored lines denote the LDS. Vertical arrows show some of the allowed optical transitions observed in the optical spectra (see Fig.~\ref{fig:OS_Er_W}).
 
 Pristine SL WS$_2$ is invariant with respect to $\sigma_h$ reflection about the z = 0 (W) plane, where the z axis is oriented perpendicular to the W plane of atoms. Therefore, electron states break down into two classes: even and odd, or symmetric and antisymmetric with respect to $\sigma_h$. $d$-orbitals of the transition metal and $p^{(t,b)}$- orbitals ($t$ and $b$ denoting the top and bottom layers) of the chalcogen atoms  give the largest contribution to the conduction and valence band structure of SL WS$_2$.\cite{SOC_agreement_1,Guinea_tight_binding_model}   Based on the $\sigma_{h}$ symmetry, the even and odd atomic orbitals are spanned by the bases $\{\phi_1=d_{x^2 - y^2}^W, {~}\phi_2=d_{xy}^W, {~}\phi_3=d_{z^2}^W,{~}\phi_{4,5}={~}p_{x,y}^{e}=(p_{x,y}^{(t)} + p_{x,y}^{(b)})/\sqrt{2},  {~}\phi_6={~}p_{z}^{e}=(p_{z}^{(t)} - p_{z}^{(b)})/\sqrt{2} \}$  and $\{\phi_7=d_{xz}^W, {~}\phi_8=d_{yz}^W, {~}\phi_{9,10}=p_{x,y}^{o}=(p_{x,y}^{(t)} - p_{x,y}^{(b)})/\sqrt{2}, {~}\phi_{11}=p_{z}^{o}=(p_{z}^{(t)} + p_{z}^{(b)})/\sqrt{2}\}$, respectively. 
 
 \begin{table}[h]%
\centering%
\begin{tabular}{ |c |c |c |c |c |c |c |}\hline
                                                  $D_{3h}$                          &    $E$    &    $\sigma_{h}$    &    $2C_{3}$    &    $2S_{3}$    &    $3C_{2}$    &    $3\sigma_{v}$\\ 
\hline
        $A'_{1}$           & 1        & 1                      & 1                 & 1                                  &            1                          &  1                        \\
 $A'_{2}$           & 1        & 1                      & 1                 &  1                                &          -1                          & -1                        \\
 $A''_{1}$           & 1        & -1                     & 1                 &  -1                              &           1                          & -1                        \\
  $A''_{2}$           & 1        & -1                     & 1                 & -1                              &          -1                          &  1                         \\
 $E'$                    & 2        & 2                      & -1                  & -1                            &         0                          &  0                        \\
 $E''$                   & 2        & -2                      & -1                  & 1                            &         0                          &  0                        \\
\hline\end{tabular}
\caption{Character table of the group $D_{3h}$. $E$, $\sigma_{h}$, $2C_{3}$, $2S_{3}$, $3C_{2}$, and $\sigma_{v}$ are the single group IRs.}
\label{table_D_3h}
\end{table}
 
\begin{figure*}
\centering
\includegraphics[width=7.2in]{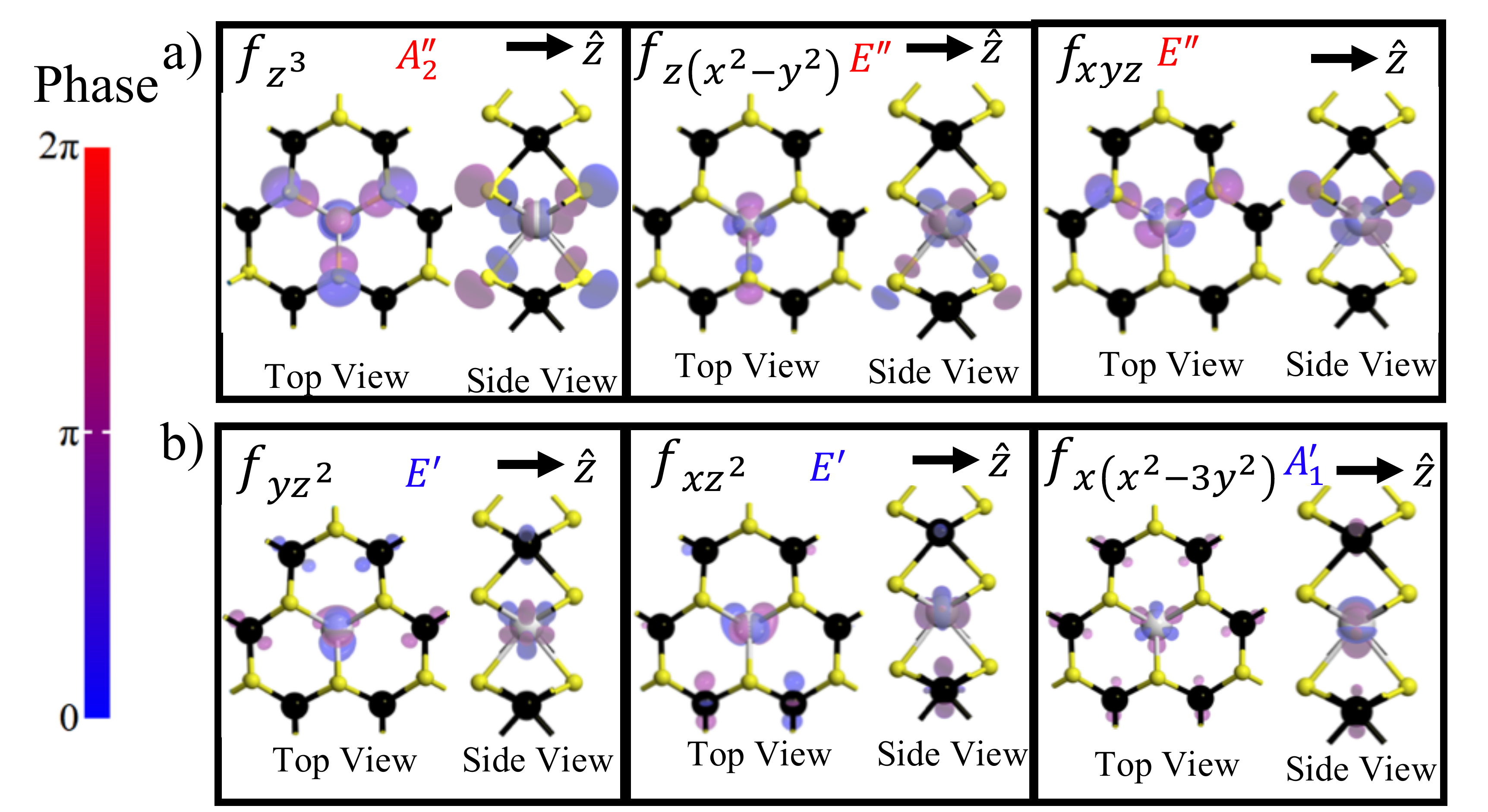}
\caption{Examples of the Bloch states for the Er$_W$ defect in $8\times8\times1$ super cell of WS$_2$. (a) Odd triplet at an energy -0.7 eV inside the valence band formed by the $f-$orbitals of the Er and $p-$orbitals of the neighboring Sulphur atoms. (b) Even triplet above the Fermi level.}
\label{fig:Bloch_states}
\end{figure*}

 The Er$_W$ defect preserves the $\sigma_{h}$ symmetry of the crystal and can be described by the group $D_{3h}$.\cite{Cheiwchammangij_SOC,Song_SOI} Symmetry operations and different irreducible representations (IRs) of D$_{3h}$ symmetry are shown in Table~\ref{table_D_3h}.  Due to $\sigma_h$ symmetry of Er$_W$ defect, the LDS break down into even and odd parity with respect to the W-plane of atoms. It can be seen in Fig.~\ref{fig:bandstructure_Er_W} and \ref{fig:Bloch_states} that LDS appear within the bandstructure as even (blue) and odd (red) states. In addition, the LDS appear in the form of triplets, i.e. a degenerate doublet and a singlet, which is in fact a consequence of 3-fold rotation symmetry $C_3$ of the defect,\cite{Erementchouk2015,Khan2017} as shown in Fig.~\ref{fig:Bloch_states}. 
 
 \subsection{Molecular Orbital Theory}
\begin{figure}[htb]
\centering
\includegraphics[width=3.5in]{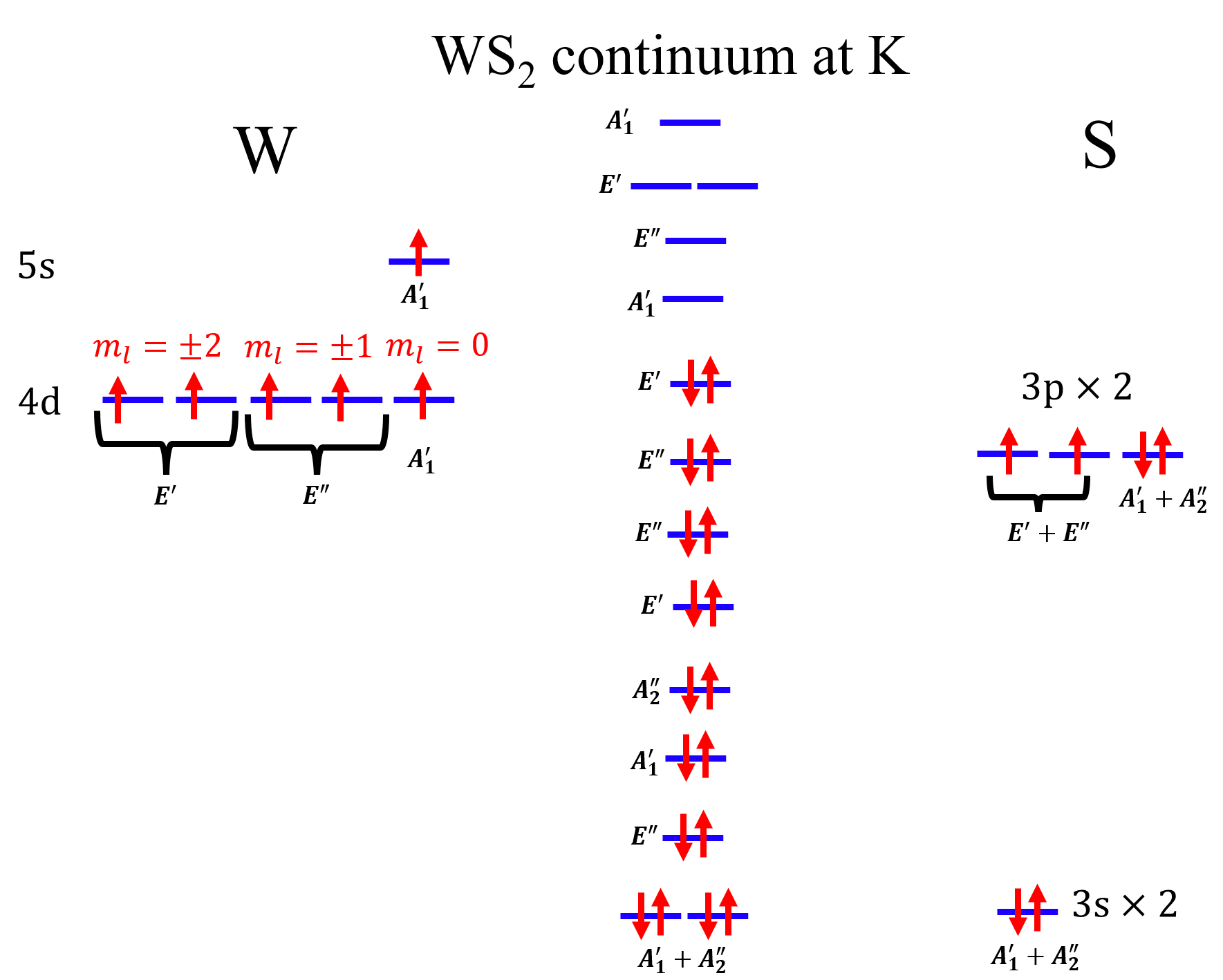}
\caption{Molecular orbital diagram of pristine WS$_2$ at the K-point. The states are labeled with the IRs of the point group $D_{3h}$.}
\label{fig:MOT_WS2}
\end{figure}

 In a simple atomistic picture a Er$_W$ defect can be regarded as an atom in an effective ligand field formed by neighboring six S atoms. In this framework the defect system decouples form the host lattice, and therefore molecular orbital theory can be applied. To analyse the various peaks that appear in the density of states (DOS), which correspond to LDS, projected density of states (PDOS) of different orbital contributions of individual atoms is presented in Fig.~\ref{fig:bandstructure_Er_W}. It can be seen that apart from the $d$ and $f$ orbitals of Er$_{\rm W}$, a substantial contribution from the $p$ orbitals of nearest neighboring S atoms and $d$ orbitals of next-nearest neighboring W atoms is present. Thus, the Hilbert base for the LDS is defined by a 23 dimensional vector, i.e. $\psi_{i}^\dagger=(\phi_{1},...,{~}\phi_{11},{~}\phi_{12}=d_{x^2 - y^2}^{Er}, {~}\phi_{13}=d_{xy}^{Er}, {~}\phi_{14}=d_{z^2}^{Er},{~}\phi_{15}=d_{xz}^{Er},{~}\phi_{16}=d_{yz}^{Er},{~}\phi_{17}=f_{z^3},{~}\phi_{18}=f_{xz^2},{~}\phi_{19}=f_{yz^2},{~}\phi_{20}=f_{xyz},{~}\phi_{21}=f_{z(x^2-y^2)},{~}\phi_{22}=f_{x(x^2-3y^2)},{~}\phi_{23}=f_{y(3x^2-y^2)})^{\dagger}$. Consequently, the electronic wave function describing a LDS state may be written as 
 \begin{equation}
     \Psi=\sum_{j}a_{j}\phi_{j},
 \end{equation} 
where $a_{i}$'s are real coefficients which can be determined from the PDOS plot, shown in Fig.~\ref{fig:bandstructure_Er_W}. Since orbitals must belong to the same IR to admix, and therefore many coefficients are zero, we can draw a molecular orbital diagram of pristine WS$_2$ as shown in Fig.~\ref{fig:MOT_WS2}. The resulting eigenstates, labeled by means of IRs of the $D_{3h}$ point group, correspond to the continuum states of the bands in WS$_2$.
The orbital energy ordering is obtained by comparison with the projected band analysis presented in Ref.~\onlinecite{Pike2017}.

\begin{figure*}[hbt]
\centering
\includegraphics[width=7in]{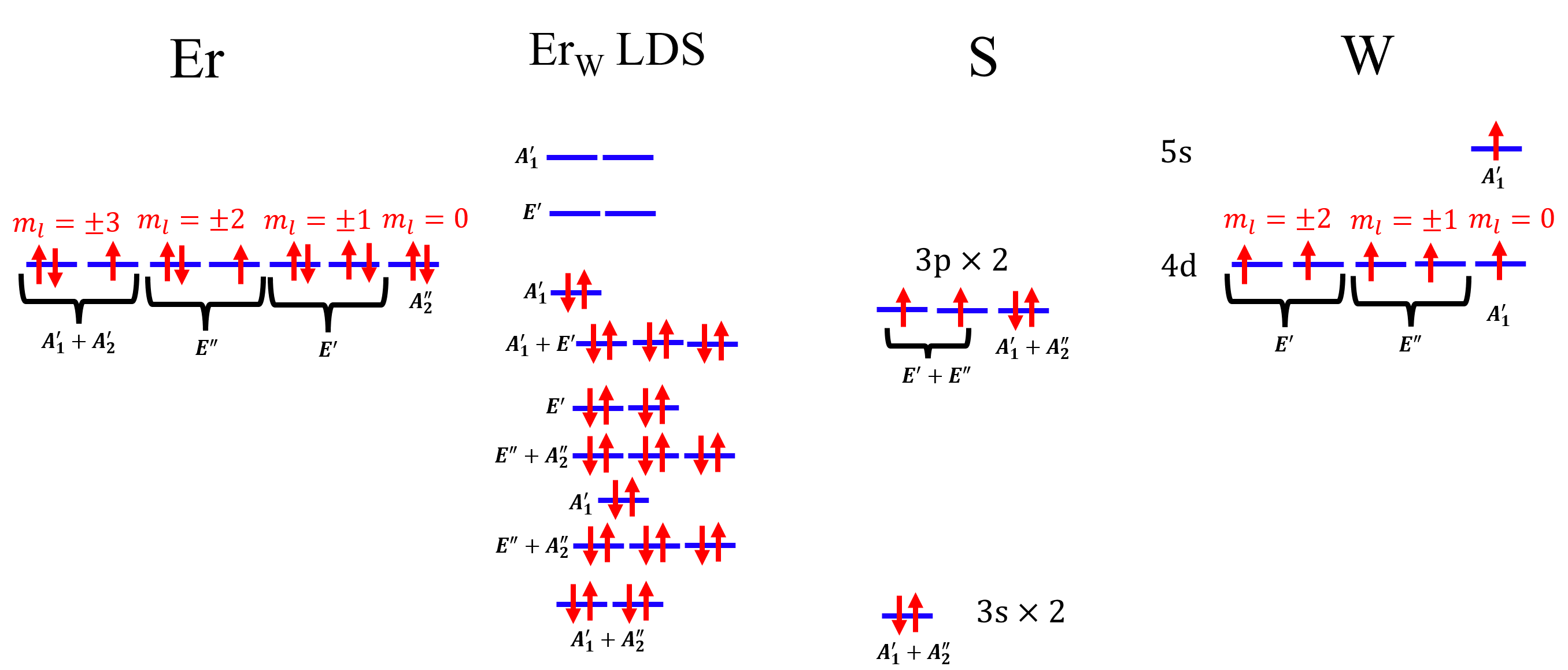}
\caption{Molecular orbital diagram of Er f orbitals in WS$_2$, giving rise to the Er$_W$ LDS shown in the bandstructure in Fig.~\ref{fig:bandstructure_Er_W}. The states are labeled with the IRs of the point group $D_{3h}$.}
\label{fig:MOT_ErW}
\end{figure*}

From the PDOS we see that the Er $f$ orbitals couple to both the $p$ orbitals of nearest neighboring S atoms and $d$ orbitals of next-nearest neighboring W atoms. Taking these two bonding types into account we can draw the molecular orbital diagram for the Er LDS, as shown in Fig.~\ref{fig:MOT_ErW}. The resulting Er LDS are labeled by means of IRs of the $D_{3h}$ point group. 
The orbital energy ordering is obtained by comparison with the PDOS shown in Fig.~\ref{fig:bandstructure_Er_W}.
We find an exact match for the highest occupied molecular orbital (HOMO) being an $A_1'$ singlet state and for the lowest unoccupied molecular orbital (LUMO) being a $E'$ doublet state, which corroborates the calculated PDOS shown in Fig.~\ref{fig:bandstructure_Er_W}.
To put this result in perspective of the size of the atoms, note that the Er atom with an average atomic radius of 1.75 {\AA} is substantially larger than a W atom with an average atomic radius of 1.35 \AA.
Therefore, it is no surprise that the resulting lattice distortions lead to relatively strong hybridizations between the $f$ orbitals of Er and the $d$ orbitals of W, which is accounted for in the bandstructure shown in Fig.~\ref{fig:bandstructure_Er_W}.

\section{Optical Response}
\label{sec:optical}

\begin{figure} 
	\begin{center}
		\includegraphics[width=3.5in]{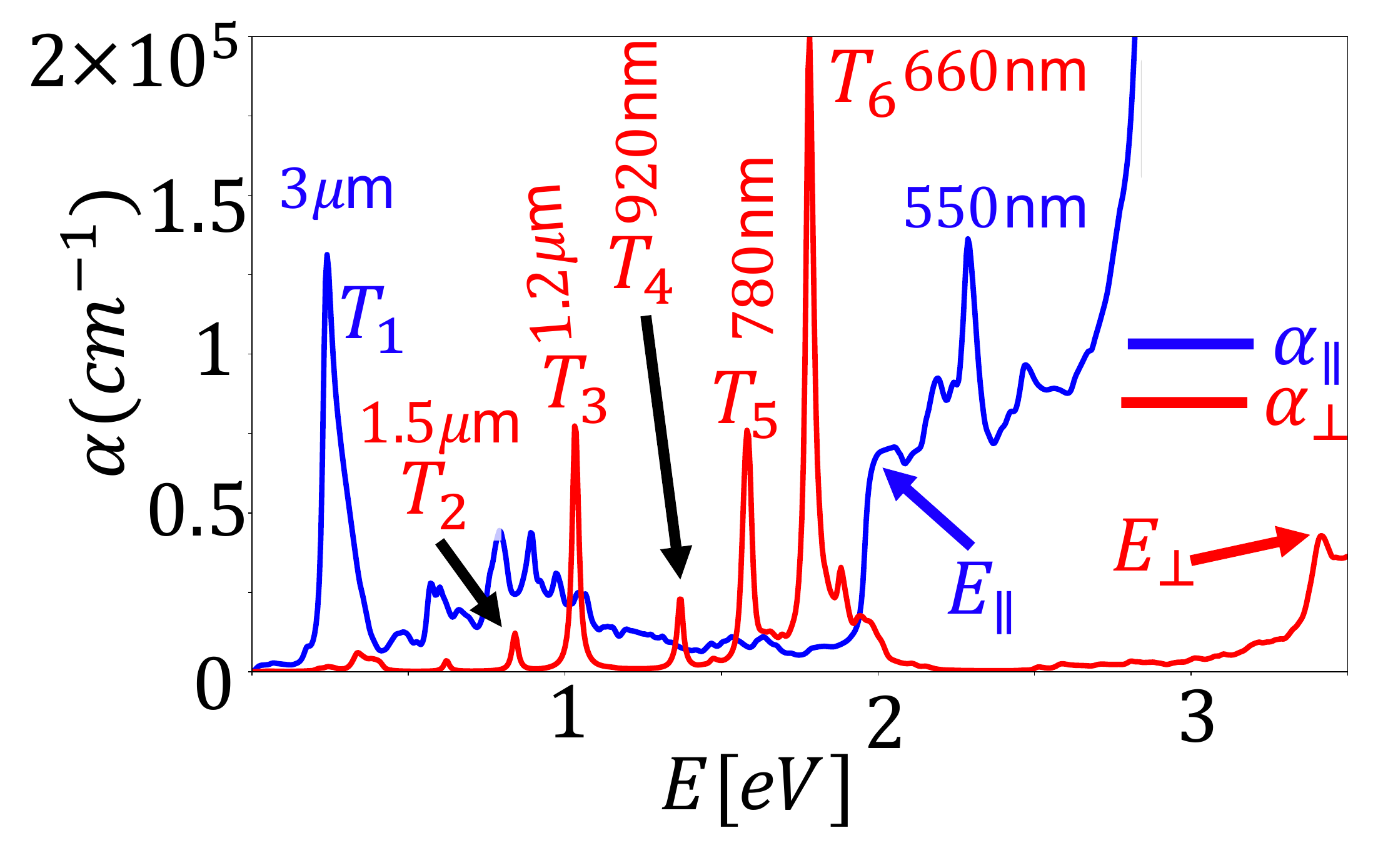}
	\end{center}
	\caption{Optical spectrum calculated by means of ATK showing resonances of Im$\chi_{\parallel}(\omega)$ (blue) and Im$\chi_{\perp}(\omega)$ (red) $a)$ due to Er$_{\rm W}$ defects in WS$_2$.}
	\label{fig:OS_Er_W}
\end{figure}

The presence of LDS in the band structure gives rise to sharp peaks in the optical spectrum. In Ref.~\onlinecite{Dielectric_function_measurement} the relative dielectric functions $\epsilon_r$ of various TMDs have been measured experimentally, which in turn are related to the absorption coefficient $\alpha$ by the standard formula $\alpha=4\pi Im[\sqrt{\epsilon_r}]/\lambda$. The dielectric tensor is evaluated using the Kubo-Greenwood formula for electric susceptibility $\epsilon_{r}-1=\chi_{ij}(\omega)$
\begin{equation}\label{eq:KGW}
\chi_{ij}(\omega)=\frac{e^{2}}{\hbar m_{e}^{2}V}\sum_{pq\bf{k}}\frac{f_{q\bf{k}}-f_{p\bf{k}}}{\omega_{pq}^2({\bf{k}})[\omega_{pq}({\bf{k}})- \omega-i\Gamma/{\hbar}]}{\bf{P}}_{pq}^{i}{\bf{P}}_{qp}^{j}
\end{equation}  
where $P_{pq}^{j}=\langle p{\bf{k}}|p^{j}|q\bf{k}\rangle$ is the $j$th component of dipole matrix element between states $p$ and $q$, $V$ the volume, $f$ the Fermi function and $\Gamma$ is the broadening, which is set to be 0.01 eV. In Fig.~\ref{fig:OS_Er_W} results for both in plane $\alpha_{\parallel}$ and out of plane $\alpha_{\perp}$ components of the absorption coefficients are presented for Er$_{\rm W}$ defects in WS$_2$. The absorption coefficient provides valuable insight into the optical selection rules for transitions between states across the Fermi level. We are interested in transitions involving states with energy near the gap edges or inside the gap. Appearance of states inside the band gap $E_{g}$ or close to the band edges leads to the resonances at single frequency $\omega_{pq}=|\omega_{p}-\omega_{q}|$. The dipole matrix element $\pi_{pq}^{j}$ determines the strength of an optical transition and whether it is allowed or prohibited by symmetries.

When considering defects in a crystal, the LDS transform according to the IRs of the symmetry group of the crystal site in which the defect resides. While the translational symmetry of the crystal is broken, point group symmetries are preserved. The Er$_{\rm W}$ defect keeps the full $D_{3h}$ symmetry. The character table for $D_{3h}$ with IRs  is shown in Table~\ref{table_D_3h}. 

The appearance of LDS inside the band gap leads to sharp resonances in $\alpha_{\parallel}$ and $\alpha_{\perp}$ at frequencies corresponding to the energy differences between LDS. However, not all transitions are allowed. Instead, several transitions are prohibited due to symmetry, i.e. when $\pi_{pq}^{j}$ does not transform according to the symmetric representation of the symmetry group of the superlattice. In the matrix element $\chi_{pq}^{j}$, the initial state $\psi_{p}$, the final state $\psi_{q}$, and the position operator $x_{j}$ transform according to the IRs $\Gamma(\psi_{p})$, $\Gamma(\psi_{q})$ and $\Gamma(x_{j})$, respectively. An electric dipole transition between two states is allowed if the direct product $\Gamma(\psi_{p})\otimes\Gamma(x_{j})\otimes\Gamma(\psi_{q})$ includes $\Gamma(I)$ in its decomposition in terms of a direct sum. $\Gamma(I)$ denotes the IR for the identity i.e., $A_{1}'$ for $D_{3h}$. This is strictly related to the polarization of the radiation. One needs to consider separately the in plane and out of plane components of ${\bf{P}}_{pq}^{j}$ because they transform according to different IRs of $D_{3h}$. The selection rules for electric dipole transitions for IRs are summarized in Table~\ref{table_D_3h_C_3v}.  We find that the out-of-plane component of ${\bf{P}}_{pq}^{j}$, which gives rise to $\pi$ transitions, is nonzero only between odd and even states of the same multiplicity (either between singlets or between doublets), while the in-plane components, which lead to $\sigma$ transitions, are nonzero for all states of the same parity.The optical transitions are in agreement with the selection rules derived in Table~\ref{table_D_3h_C_3v}.

\begin{table}[h]
\begin{tabular}{|c |c |c |c |c |c |c|}\hline
$D_{3h}$ & $A_{1}^{\prime}$ & $A_{2}^{\prime}$ & $A_{1}^{\prime\prime}$ & $A_{2}^{\prime\prime}$ & $E^{\prime}$ & $E^{\prime\prime}$   \\
\hline
$A_{1}^{\prime}$ & & &    & $\pi$      &  $\sigma$  &   \\
\hline
 $A_{2}^{\prime}$  &  &     & $\pi$   &               & $\sigma$ &     \\
\hline
$A_{1}^{\prime\prime}$  &      & $\pi $  & & &     & $\sigma$      \\
\hline
$A_{2}^{\prime\prime}$ &  $\pi$ & & & & & $\sigma$\\
\hline
$E^{\prime}$ & $\sigma$&$\sigma$ & & & $\sigma$&$\pi$\\
\hline
$E^{\prime\prime}$ & & & $\sigma$&$\sigma$ & $\pi$&$\sigma$\\
\hline
\end{tabular}
\caption{Electric Dipole selection rules in $D_{3h}$ symmetry. $\sigma$ represents in plane transitions while $\pi$ represents out of plane transitions.}
\label{table_D_3h_C_3v}
\end{table}

It can be seen that some of the optical transitions are in reasonably good agreement with available experimental data for optical transitions in Er$^{3+}$:YAG.
The comparison for the resonances wavelength of the optical emissions between Er$^{3+}$:YAG and Er$^{3+}$:WS$_2$  is shown in Table~\ref{tab:comparison}. The labeling of the Er LDS is only valid for Er$^{3+}$:YAG, but might reveal insight for future studies for identifying the LDS in Er$^{3+}$:WS$_2$.
There is also reasonable to very good agreement between the resonance wavelengths we obtain from the numerical calculations and the resonance wavelengths of 1.5 $\mu$m, 980 nm, and 800 nm observed experimentally in Er:MoS$_2$ atomic layers and Yb/Er:WSe$_2$.\cite{Bai2016, Bai2018} 

\begin{table}[h]
\begin{tabular}{|c ||c |c |}\hline
Er$^{3+}$ states & Er$^{3+}$:YAG & Er$^{3+}$:WS$_2$   \\
\hline\hline
$^4$I$_{11/2}\rightarrow$ $^4$I$_{13/2}$  & 3 $\mu$m &  3 $\mu$m ($T_1$) \\
\hline
$^4$I$_{13/2}\rightarrow$ $^4$I$_{15/2}$  & 1.5 $\mu$m &  1.5 $\mu$m ($T_2$)  \\
\hline
$^4$I$_{11/2}\rightarrow$ $^4$I$_{15/2}$  & 965 nm &  1.2 $\mu$m ($T_3$)  \\
\hline
$^4$I$_{9/2}\rightarrow$ $^4$I$_{15/2}$  & 860 nm &  920 nm ($T_4$) or 780 nm ($T_5$)  \\
\hline
$^4$F$_{9/2}\rightarrow$ $^4$I$_{15/2}$  & 677 nm &  660 nm ($T_6$)  \\
\hline
$^4$F$_{7/2}\rightarrow$ $^4$I$_{15/2}$  & 545 nm &  550 nm  \\
\hline
\end{tabular}
\caption{Resonance wavelength of optical transitions (absorption and emission) for Er$^{3+}$:YAG and Er$^{3+}$:WS$_2$. The absorption transitions in Er$^{3+}$:WS$_2$ are labeled according the optical spectrum shown in Fig.~\ref{fig:OS_Er_W}.}
\label{tab:comparison}
\end{table}

\section{Conclusion}
In this paper we have provided numerical and analytical descriptions of electronic and optical properties of SL WS$_2$ in the presence of Er$_{\rm W}$ defects. We have shown that the LDS related to Er$_W$ defect gives rise to sharp transitions both in $\alpha{\parallel}$ and $\alpha_{\perp}$. In order to understand defect related optical transitions, odd states, which are usually neglected for pristine cases, need to be considered in addition to even states. A central result of our paper is that group theory can be used to derive strict selection rules for the optical transitions, which are in excellent agreement with the absorption spectrum calculated using the Kubo-Greenwood formula using the Kohn-Sham orbitals. 

Given the atomically sharp optical transition lines  at 3 $\mu$m and 1.5 $\mu$m, the decoherence time should be very long, which would make Er$^{3+}$:WS$_2$ a promising candidate for single-photon emitter for quantum communication and qubit for quantum computing. 
There are several advantages for using  Er$^{3+}$:WS$_2$ in a quantum platform over other currently existing solid-state qubits.
One of the main problems for implementing large-scale quantum networks and quantum computing is decoherence. With respect to that, TMDs have several major advantages: 
\begin{itemize}
\item W has a weak abundance of 14\% of nuclear spin ½ (183W) and S has a negligibly small abundance of 0.8\% of nuclear spin 3/2 (33S), in stark contrast to the NV centers in diamond, where N has a 99.6\% abundance of nuclear spin 1 (14N) and C has a 20\% abundance of nuclear spin 3/2 (9C and 11C) and a 15\% abundance of nuclear spin ½ (15C). Consequently, there will be a much weaker decoherence of the Er spin state due to the environmental nuclear spins, given also the fact that Er substitutes W and thus Er binds covalently to the surrounding S atoms, which are nearly free of nuclear spins.
\item The location of the defects is accurate on the atomic length scale in z-direction, perpendicular to the plane of the 2D material, in contrast to 3D materials, such as diamond. This will result in a much higher sensitivity for quantum sensing due to accurate distance to target atoms.
\item 2D materials have clean surfaces, in stark contrast to diamond that hosts dark P1 nitrogen defects with nuclear spins that also lead to decoherence of the spin state of the NV center.
\end{itemize}

These advantages suggest that REIs embedded in 2D materials made of TMDs might be vastly superior to NV centers in diamond and could pave the way to realizing scalable quantum networks, scalable quantum computing, and ultrasensitive remote quantum sensing. 

During the preparation of this manuscript we learned of similar ab-initio calculations for electronic and optical properties of Er$_{\rm W}$ defects in WS$_2$ using VASP.\cite{Lopez2021}
While VASP is based on plane-wave basis sets, ATK is based on atomic orbital basis sets. That is why there are differences in the results.

\begin{acknowledgments}
We acknowledge support provided by the Airforce Summer Faculty Fellowship 2021. We acknowledge useful discussions with Carlos A. Meriles, Johannes Flick, and Gabriel I. L\'opez-Morales. 
\end{acknowledgments}


\section{Author's Contributions}
All authors contributed equally to this work.

\section{Data Availability}
The data that support the findings of this study are available from the corresponding author upon reasonable request.

\section{References}

\end{document}